\documentclass[a4paper,11pt]{article}
\usepackage{jcappub} 
\usepackage{lineno}
\usepackage{siunitx}
\usepackage{rotating}
\usepackage{multirow}
\usepackage{makecell}
\usepackage{array}
\usepackage{sidecap}
\usepackage{tabularx}
\usepackage{amsmath}

\title{\boldmath The muon charge asymmetry and the directional distribution of thunderstorm events observed by the GRAPES-3 muon telescope}

\author[a,1]{B. Hariharan\note{Corresponding author}}
\author[a]{S.K. Gupta}
\author[b]{Y. Hayashi}
\author[a]{P. Jagadeesan}
\author[a]{A. Jain}
\author[b]{S. Kawakami}
\author[c]{H. Kojima}
\author[a]{P.K. Mohanty}
\author[d]{Y. Muraki}
\author[a]{P.K. Nayak}
\author[c]{A. Oshima}
\author[a]{M. Rameez}
\author[a]{K. Ramesh}
\author[a]{L.V. Reddy}
\author[c]{S. Shibata}

\affiliation[a]{Tata Institute of Fundamental Research, Homi Bhabha Road, Mumbai, 400005, India}
\affiliation[b]{Graduate School of Science, Osaka City University, Osaka, 558-8585, Japan}
\affiliation[c]{College of Engineering, Chubu University, Kasugai, 487-8501, Japan}
\affiliation[d]{Institute for Space-Earth Environmental Research, Nagoya University, Nagoya, 464-8601, Japan}

\emailAdd{89hariharan@gmail.com}

\abstract{The electric fields inside thunderstorms can significantly modify the intensity of secondary cosmic ray muons at the ground level, producing measurable variations in their intensity (\imu{}). By utilizing the decade-long observations of thunderstorms (April 2011--December 2020) by the GRAPES-3 muon telescope (G3MT), a directional asymmetry in \imu{} is observed, with nearly six times more events being detected from the east than the west directions. Using detailed CORSIKA Monte Carlo simulations, it is shown that this asymmetry is caused by the variations of the muon charge ratio  R$_\mu$ (N$_{\mu^+}$/N$_{\mu^-}$). The anisotropic R$_\mu$ in turn, is caused by the systematic changes in geomagnetic cutoff rigidities, and subsequent selective filtering of predominantly positively charged primary cosmic rays. As a consequence, the R$_\mu$ increases systematically from west to east across the G3MT field of view, enhancing the sensitivity of east directions to positively charged thunderstorm top. Monte Carlo simulations with constant R$_\mu$ show that the directional asymmetry disappears, demonstrating the muon charge imbalance to be the dominant driver of the observed asymmetry. The dependence of R$_\mu$ on the hadronic interaction is also studied by comparing seven combinations high-, and low-energy hadronic interaction generators, which show a $\lesssim$7\% spread in R$_\mu$, and $\lesssim$14\% variation in the derived thunderstorm potentials. These results provide the first quantitative link between the muon charge asymmetry caused by the geomagnetic field, and the directional distribution of thunderstorms, reinforcing the role of muon observations as a probe of gigavolt potentials in atmospheric electrical structures.}

\newcommand{\imu}{$\Delta$I$_{\mu}$}
\date{}
\begin{document}
\maketitle
\flushbottom

\section{Introduction} 

The study of primary cosmic ray (PCR) interactions in the atmosphere has long provided a window into both the high-energy interactions, as well as the atmospheric electrodynamics. The PCRs undergo interactions after entering the atmosphere, producing cascades of secondary mesons that eventually decay into muons, and neutrinos \cite{Gaisser_2016,Grieder_2010,Sanuki_2007}. Because the muons are highly penetrating particles and largely retain the directional memory of their parent mesons, their intensity and charge composition at the surface carries signatures of PCR composition, and that of the environmental influences \cite{Allkofer_1978,Burnett_1973}. The muon charge ratio R$_\mu$, defined as the ratio of numbers of the positive (N$_{\mu^+}$), and negative muons (N$_{\mu^-}$) is a sensitive observable encapsulating the production, and propagation of the charged secondaries. Early ground-based measurements reported R$_\mu$\,=\,1.26 in the multi-GeV range, while the later precision measurements by MUTRON \cite{Muraki_1979,Muraki_1983}, CMS \cite{CMS}, and MINOS \cite{MINOS_2011,MINOS_2016} experiments confirmed this ratio, and its mild energy dependence up to the TeV scale. The underground, and mountain based experiments have also shown that R$_\mu$ depends on the zenith angle, and the geomagnetic latitude \cite{AlvarezCompton_1933,Barber_1949}. The directional dependence of muon intensity, commonly known as the ``east–west'' effect, was first discovered in the 1930s \cite{AlvarezCompton_1933,Johnson_1933_1,Johnson_1933_2,Johnson_1935,Johnson_1941,Barber_1949,OgawaNagahara_1950,Rossi_1964,Lipari_2000}. This asymmetry originates from the geomagnetic field, which differentially filters PCRs according to their charge sign, and rigidity. Predominantly, positively charged PCRs from the east experience stronger magnetic deflection than those from the west, resulting in a lower flux from the east due to higher cutoff rigidities \cite{Mendonca_2019}. Subsequent studies demonstrated that this geomagnetic filtering translates into measurable asymmetries in the muon intensity, and their charge ratio at the ground level \cite{Rebel_2007}. At the geomagnetic latitude of Ooty (11.4$^\circ$N), where the cutoff rigidity ranges from $\sim$15\,GV in the west to $\sim$24\,GV in the east \cite{Hariharan_2019_2} -- such directional effects cause significant variations in R$_\mu$.

In addition to the geomagnetic effects, the atmospheric electric fields produced inside major thunderstorms also influence the muon intensity. Early balloon, and aircraft experiments detected vertical electric fields of several hundred kV/m, and potentials of $\sim$100 million volts \cite{Marshall_1991,Dwyer_2009}. More recently, the GRAPES-3 experiment provided the first direct evidence for gigavolt potential in a thunderstorm by correlating short-term variations in muon intensity with simultaneous measurements of electric field on the ground \cite{Hariharan_2019_1,Hariharan_2025}. The underlying physics involves either acceleration or deceleration of the charged muons traversing the thunderstorm region, which depends on the polarity of the muon, and the orientation of electric field inside the thunderstorm, leading to either excess or deficit in the ground-level muon intensity \cite{Alexeenko_1985,Alexeenko_1987,Alexeenko_2001,Alexeenko_2002}. This mechanism forms the basis of interpretation of thunderstorm induced muon intensity variations (\imu{}) observed by the GRAPES-3 muon telescope (G3MT) \cite{Hariharan_2019_1,Hariharan_2025}, Mt. Norikura \cite{Muraki_2004}, Telescope Array \cite{Abbasi_2022}, HAWC \cite{Jimenez_2019,Bowers_2019}, LHASSO \cite{Yan_2020}, etc. However, a highly directional asymmetry in the frequency of thunderstorms is observed in the G3MT data, wherein number of events recorded in the east are six times larger than in the west directions. Explanation of this directional asymmetry is obtained with the aid of CORSIKA Monte Carlo simulations \cite{corsika}. In the CORSIKA package different combinations of high-, and low-energy hadronic interaction generators are used to simulate the muon intensity throughout the atmosphere. These simulations show that the value of R$_\mu$ in the G3MT field of view (FOV) varies, systematically increasing from the west ($\approx$1.14) to the east ($\approx$1.37) direction. The regions with a higher R$_\mu$ display larger variations in the muon intensity for the same thunderstorm potential resulting in a larger number of events from the east than the west directions.

The interplay of hadronic physics, geomagnetic effects, and thunderstorm electric fields has been explored in several previous theoretical work \cite{Buitink_2010,Bektasoglu_2012,Dorman_2003}, but the explicit link between the muon charge asymmetry and the directional dependence of detected thunderstorms has not been quantitatively investigated. Recent GRAPES-3 analyses demonstrated that the inferred thunderstorm potential depends weakly ($\lesssim$14\%) on the choice of hadronic interaction generators \cite{Hariharan_2025}. This highlights the need for improved modeling of muon production, which influences the detection of thunderstorm by G3MT. Results from other major experiments provide evidence that the thunderstorm electric fields significantly impact the intensity of charged secondaries including muons \cite{Muraki_2004,Abbasi_2022,Jimenez_2019,Bowers_2019,Yan_2020}. The present work builds on the body of existing work by investigating the factors affecting R$_\mu$, which in turn influences the directional distribution of thunderstorm events. In the present study, 487 events recorded by the G3MT between April 2011 and December 2020 are analyzed. As mentioned above, these data show a strong directional asymmetry, with about six times more events being observed from the east than the west directions. By using CORSIKA Monte Carlo simulations, the directional dependence of R$_\mu$ is obtained, and compared with the observed thunderstorm data. Among the hadronic interaction generators used in CORSIKA, the SIBYLL–FLUKA combination had yielded the highest muon charge asymmetry, and therefore, the most conservative estimate of electric potentials. All combinations predict the same qualitative behavior, and the magnitude of R$_\mu$ varies by up to 7\% between different generator sets. However, the underlying trend has remained unchanged, namely, the muon charge asymmetry being the primary cause for asymmetric angular distribution of thunderstorms. This result has strengthened the case for using the muon observations to study massive thunderstorm potentials, and their interactions with the geomagnetic field. It also reinforces the utility of G3MT as a unique tool for probing the gigavolt-scale electric structures in the thunderstorms.

In the following, the Section\,\ref{G3MT} provides a brief account of the G3MT and its capabilities, followed by Section\,\ref{Events}, which describes the statistics of thunderstorm events recorded from April 2011 to December 2020. Section\,\ref{Impact} provides a detailed account of the influence of muon charge asymmetry on the number of thunderstorms recorded by G3MT, with its subsections further detailing the physics of muon charge asymmetry (Section\,\ref{Asymmetry}), and Monte Carlo simulations (Section\,\ref{MC}) carried out to explain the influence of muon charge asymmetry on distribution of thunderstorm events. Section\,\ref{Influence} explores the dependence of the muon charge asymmetry and its impact on the angular distribution of thunderstorm events on the choice of high-, and low-energy hadronic interaction generators. This work is summarized and conclusions presented in Sections \ref{Discussion}, and \ref{Conclusions}, respectively.

\section{The GRAPES-3 muon telescope (G3MT) \label{G3MT}}

\begin{figure}[t]
    \centering
    \includegraphics[width=0.99\textwidth]{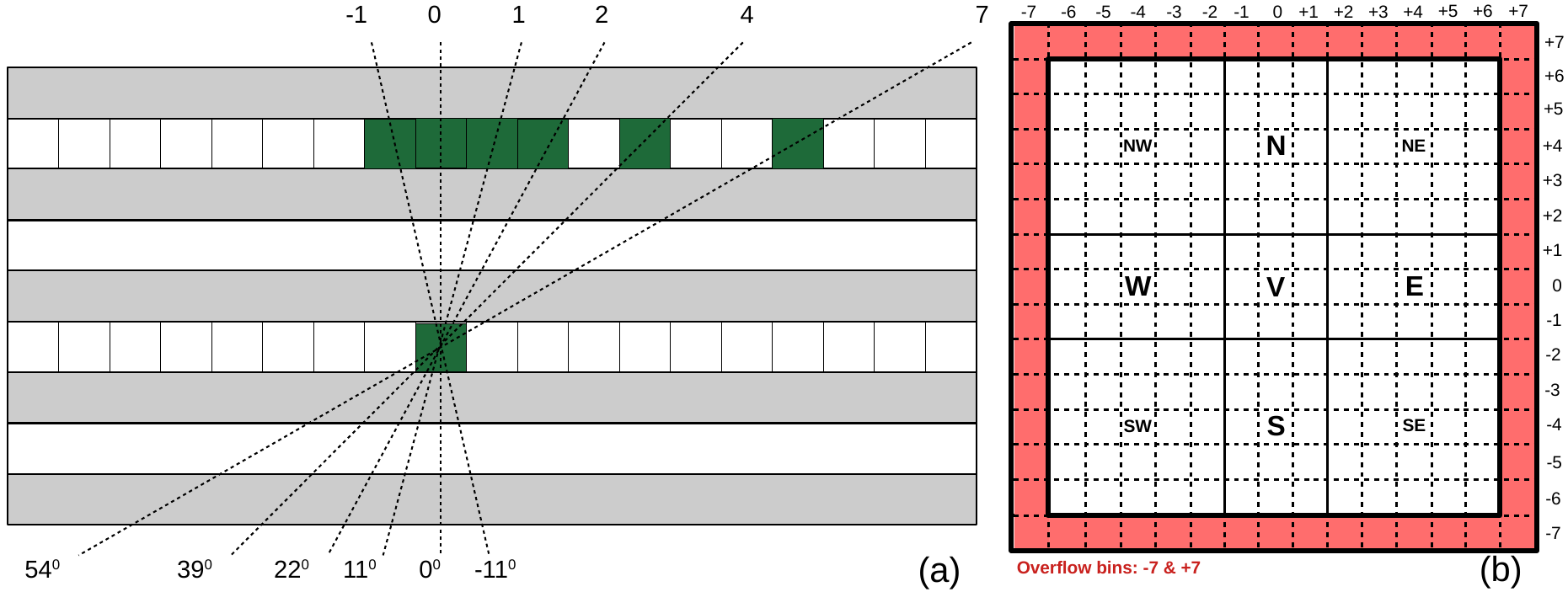}
    \caption{Representation of (a) muon angle reconstruction in a single projection using PRC hits, and (b) solid angle coverage of G3MT in 169-direction configuration shown as dotted lines, excluding outermost overflow bins. Solid lines represent coarser 9-direction configuration.}
    \label{Fig_MA}
\end{figure}

G3MT is one of the two main components of the GRAPES-3 experiment located in Ooty, India (11.4$^\circ$N, 76.7$^\circ$E, 2200\,m above mean sea level). The G3MT is built from 3712 proportional counters (PRCs) distributed into sixteen independent muon modules. The PRCs are made of sealed mild steel tubes of wall thickness 2.3\,mm. The dimensions of each PRC are 600$\times$10$\times$10\,cm$^3$, and they are filled with P10 gas (a mixture of 90\% argon, and 10\% methane). The PRCs are operated at about +3000\,V$_{DC}$ to ensure sufficient charge multiplication for the detection of singly charged particles. Each muon module has four PRC layers separated by 15\,cm concrete slabs. In each projection place, the spatial separation of the PRCs hit in the upper, and lower layers are calculated as shown in Figure\,\ref{Fig_MA}a (vertical muon track defines the direction 0). The inclined muon tracks are selected up to a separation of --7 to +7 PRCs, for a total of 15 directions in each projection plane. The muons tracks inclined beyond the 7$^{th}$ PRC are also stored in the direction 7, which consequently also becomes the overflow direction for the respective plane. A 2\,m thick layer of concrete blocks placed on the top of each module with 45$^\circ$ inclined inverted pyramidal design provides an energy threshold of sec($\theta$)\,GeV for muons incident at a zenith angle of $\theta$. All sixteen muon modules collectively span an area of 560\,m$^2$. This geometrical arrangement of the G3MT provides a 2.3\,sr coverage of the overhead sky with a mean angular resolution of about 4$^\circ$. By combining the information from the two projections, the direction of each muon is binned into one of the 225 directions (15$\times$15) as shown in Figure\,\ref{Fig_MA}b. However, for further data analysis, the overflow directions are not used, therefore only the information from the inner 13$\times$13\,=\,169 directions are used. The G3MT records about four billion muons every day in the total FOV. The inner grid of 169 directions can be combined into coarser directions according to the requirements of the physics analysis. The 9-direction configuration has been widely used for most of the GRAPES-3 studies. More details about the G3MT may be found in \cite{Hayashi_2005}.
     
The second component of GRAPES-3 is an array of 400 plastic scintillator detectors (G3SD) that covers a total area of 25000\,m$^2$. Each detector with an area of 1\,m$^2$, records the relative arrival time, and the energy deposited by the secondary particles in an extensive air shower (EAS) through the trigger generated by the G3SD. These measurements allow reconstruction of the core location, age, and the size of each EAS. The G3SD records about 3.5 million EASs produced by PCRs in the energy range 10\,TeV--10\,PeV every day. A detailed account of the G3SD is given elsewhere \cite{Gupta_2005}. The G3MT is equipped with two distinct data acquisition systems (DAQs). The first DAQ records the PRC hits for each passing muon with reference to the EAS trigger. An offline reconstruction allows determination of the total number of muons detected for each EAS trigger. Subsequently, the nuclear mass composition of the PCRs can be derived by using these muon measurements. The second DAQ records the muon tracks reconstructed online independent of the EAS trigger. The GeV muons recorded by the G3MT are an excellent proxy for the study of thunderstorms, geomagnetic storms, and other exotic phenomena \cite{Hariharan_2019_1,Mohanty_2016,Hariharan_2023}.

\section{Thunderstorm events observed by G3MT \label{Events}}

\begin{figure}[t]
    \centering
    \includegraphics[scale=0.50]{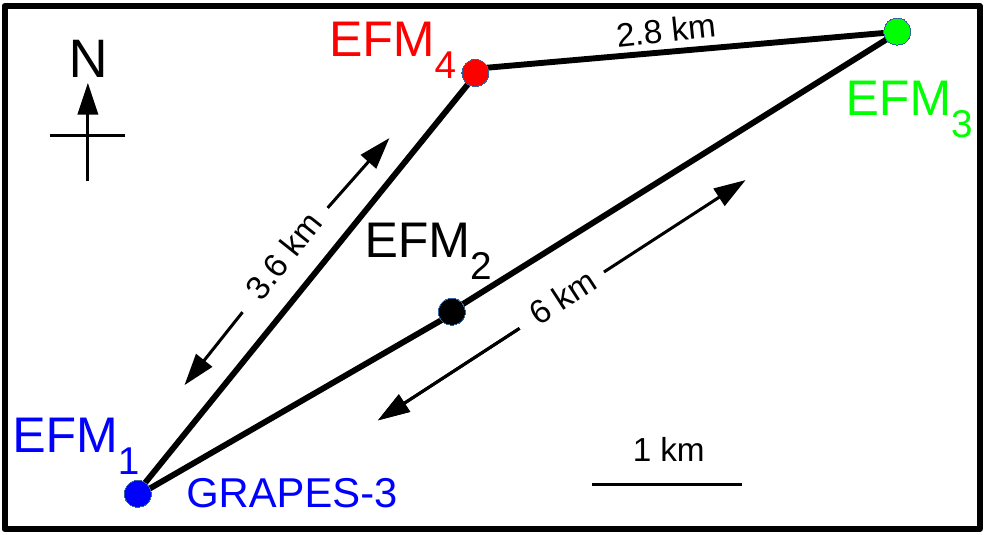}
    \caption{Locations of electric field mills installed in GRAPES-3 and other places, labeled 1 to 4.}
    \label{Fig_EFM}
\end{figure}

\begin{figure}[t]
    \centering
    \includegraphics[scale=0.45]{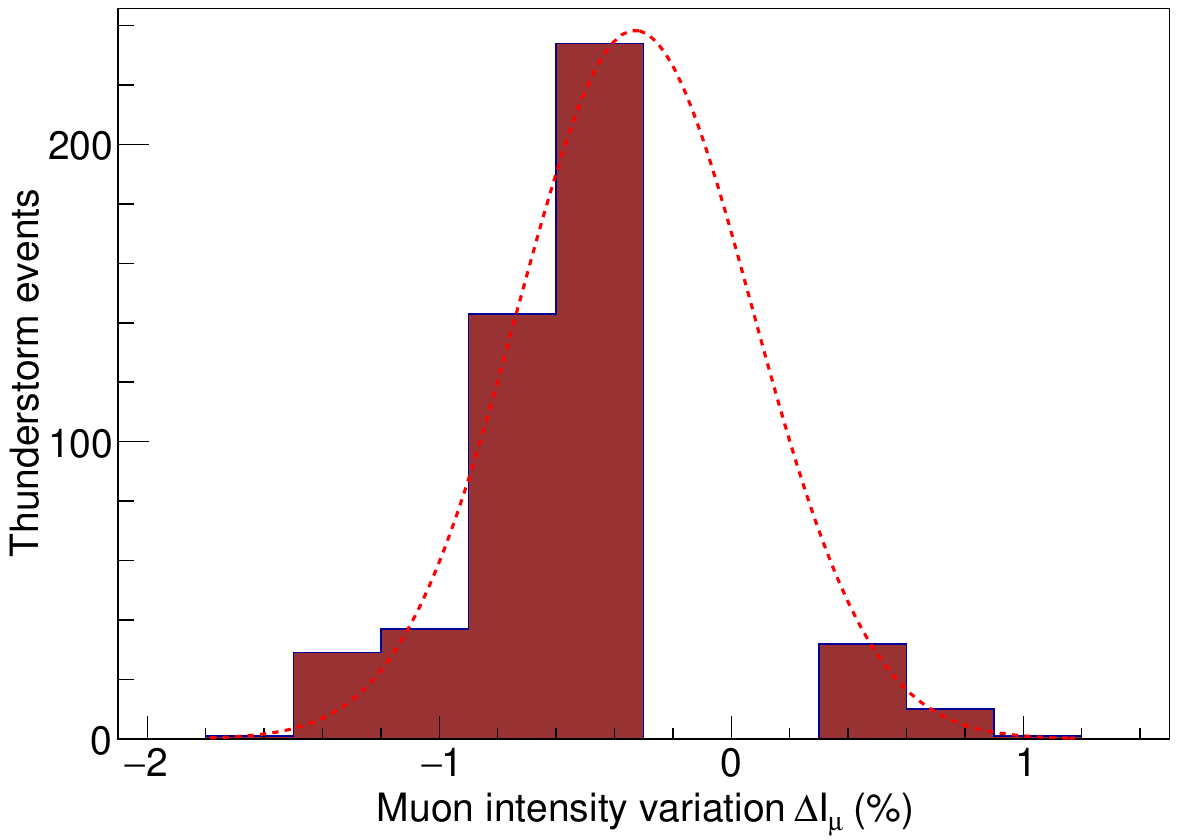}
    \caption{Distribution of 487 thunderstorm recorded by G3MT during April 2011 to December 2020 based on observed muon intensity variation ($\Delta I_{\mu}$). Since events are selected based on condition $\mid\Delta I_{\mu}\mid>$0.3\%, bins in band --0.3 to 0.3\% are empty. Dotted line is a Gaussian fit to guide the eye, and indicate that event distribution follows a normal distribution. The fits yields a mean\,=\,--0.33\%, and rms\,=\,0.4\% -- negative mean implies that most thunderstorms cause muon deficit due to dominant positive thunderstorm potential in the nature.}
    \label{Fig_Events}
\end{figure}

The atmospheric electric field (V/m) plays an important role in the study of thunderstorm properties. Even during fair weather conditions, a vertical electric field of $\sim$120\,V\,m$^{-1}$ exists near the ground, which is produced by the global electrical circuit \cite{Rycroft08}. An electric field mill (EFM) is a robust, and sensitive device that is widely used for monitoring the local electric fields \cite{Antunes_2020}. Four ``Boltek EFM-100'' EFMs \cite{Boltek} were installed in the vicinity of the GRAPES-3 to continuously monitor the atmospheric electric field. The deployment of the four EFMs is schematically shown in Figure\,\ref{Fig_EFM}. The EFMs have been in continuous operation since April 2011, recording the local electrical field with a resolution of 0.01\,V/m every 50\,milliseconds. The electric field data from the EFMs when combined with the muon measurements by the G3MT allow real-time tracking of the movement of a thunderstorm including its altitude, speed, area, and (giga\,volt) electric potential \cite{Hariharan_2019_1}. However, in case of most of the thunderstorms, the time overlap of several thunderstorm events results in extremely complex electric field profiles, which render the task of estimating their properties nearly impossible. In a recent study of the decade-long G3MT dataset containing 487 thunderstorms, eight major events with gigavolt potentials were identified, which underscores the fact that generation of such potentials is not a very uncommon feature of thunderstorms \cite{Hariharan_2025}.

\renewcommand{\arraystretch}{1.4}
\begin{table}[t]
    \centering
    \begin{tabular}{|c|c|c|c|}
        \cline{1-3}
        (NW)          & (N)           & (NE)          & \multicolumn{1}{c}{\textbf{(N$_G$)}}   \\
        6.2\%         & 1.8\%         & 30.1\%        & \multicolumn{1}{c}{\textbf{38.1\%}} \\
        1.13$\pm$0.05 & 1.26$\pm$0.03 & 1.38$\pm$0.06 & \multicolumn{1}{c}{\textbf{1.26}}   \\
        \cline{1-3}
        (W)           & (V)           & (E)           & \multicolumn{1}{c}{\textbf{(H$_G$)}}   \\
        0.6\%         & 0.2\%         & 2.8\%         & \multicolumn{1}{c}{\textbf{3.6\%}}  \\
        1.14$\pm$0.05 & 1.24$\pm$0.02 & 1.35$\pm$0.06 & \multicolumn{1}{c}{\textbf{1.24}}   \\
        \cline{1-3}
        (SW)          & (S)           & (SE)          & \multicolumn{1}{c}{\textbf{(S$_G$)}}   \\
        6.9\%         & 2.8\%         & 48.6\%        & \multicolumn{1}{c}{\textbf{58.3\%}} \\
        1.13$\pm$0.05 & 1.26$\pm$0.03 & 1.39$\pm$0.06 & \multicolumn{1}{c}{\textbf{1.26}}   \\
        \cline{1-3}
        \multicolumn{1}{c}{\textbf{(W$_G$)}}   & \multicolumn{1}{c}{\textbf{(V$_G$)}}  & \multicolumn{1}{c}{\textbf{(E$_G$)}} 
        \\
        \multicolumn{1}{c}{\textbf{13.7\%}} & \multicolumn{1}{c}{\textbf{4.8\%}} & \multicolumn{1}{c}{\textbf{81.5\%}} \\
        \multicolumn{1}{c}{\textbf{1.14}}   & \multicolumn{1}{c}{\textbf{1.25}}  & \multicolumn{1}{c}{\textbf{1.37}}
    \end{tabular}
    \caption{Percent of thunderstorm events in 9-direction configuration of GRAPES-3 field of view shown inside the cells of table, labeled by the directions, namely, north-west (NW), north (N), north-east (NE), west (W), vertical (V), east (E), south-west (SW), south (S), and south-east (SE). Table contains 487 thunderstorm events observed during April 2011--December 2020. The rightmost and bottommost values outside the table display total events in the respective directional groups along horizontal and vertical axes, northern (N$_G$), horizontal (H$_G$), and southern (S$_G$) and western (W$_G$), vertical (V$_G$), and eastern (E$_G$) groups, respectively.}
    \label{Tab_1}
\end{table}

\begin{figure}[t]
    \centering
    \includegraphics[scale=0.60]{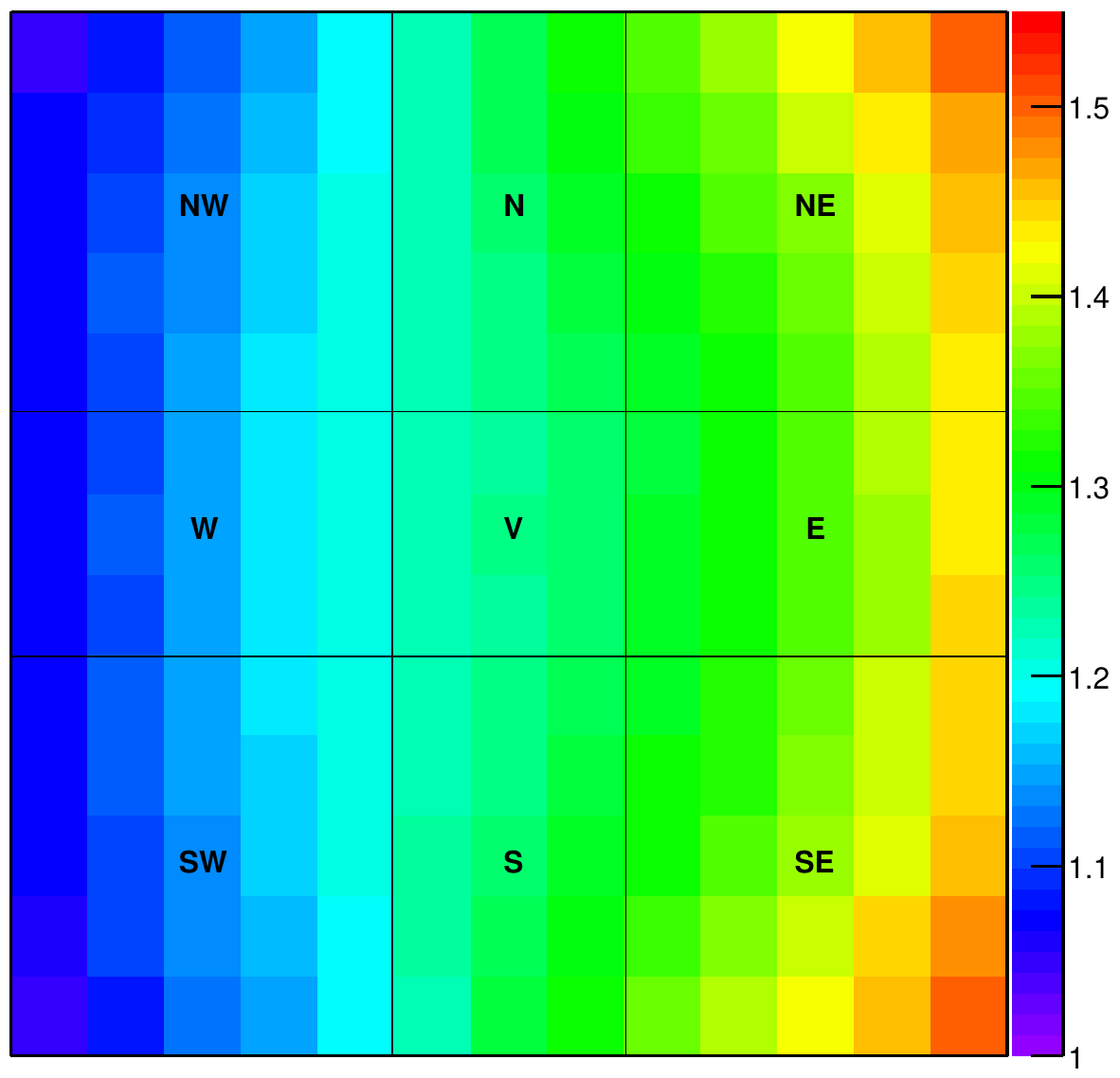}
    \caption{Map of muon charge ratio R$_\mu$ (N$_{\mu^+}$/N$_{\mu^-}$) derived by using CORSIKA with SIBYLL-FLUKA combination in 169-direction configuration of G3MT FOV. A sample of 10$^6$ simulated muons yields a statistical error of $\sim$0.1\%. R$_\mu$\,=\,1.251 for full FOV of G3MT, and has values of  1.136, 1.254, 1.374 for vertical groups in west (W$_{G}$), vertical (V$_{G}$), east (E$_{G}$), respectively. Similarly, R$_\mu$ has values of 1.258, 1.245, 1.261 for horizontal groups in north (N$_{G}$), horizontal (H$_{G}$), south (S$_{G}$), respectively. The mean values of R$_\mu$ of 1.255$\pm$0.119, and 1.255$\pm$0.008 for vertical, and horizontal groups, respectively are nearly same but the large difference in their rms indicates presence of a stronger muon charge asymmetry in E-W (rms\,=\,0.119) than N-S (rms\,=\,0.008) directions.}
    \label{Fig_Ratio_Map}
\end{figure}

Although the G3MT is operating continuously for more than two decades, the analysis of thunderstorms became possible only after the installation of EFMs in April 2011. Based on a muon intensity cut of $\mid\Delta I_{\mu}\mid>$0.3\%, a total of 487 thunderstorms were identified during April 2011--December 2020, implying a detection rate of $\sim$50 events/year. The distribution of these events is shown in Figure\,\ref{Fig_Events}, which seems to follow a normal distribution with mean\,=\,--0.33\%. The negative mean value is reflective of the fact that most events produce a muon deficit. This in turn is the consequence of the predominant occurrence of positive thunderstorm potentials in thunderstorms. These events are further analyzed along 9-directions of G3MT, each of which has a rather similar FOV, as shown in Figure\,\ref{Fig_MA}b. These directions are labeled north-west (NW), north (N), north-east (NE), west (W), vertical (V), east (E), south-west (SW), south (S), and south-east (SE) -- hereafter, for brevity only the labels in parentheses would be used. In Table\,\ref{Tab_1}, percent of thunderstorm events in each of the nine directions are listed. This highly asymmetric distribution shows that the largest percent of events occurred in SE (49\%), and another 30\% in the NE direction. Together, the three east directions account for majority of events (82\%), the remaining six directions recorded only $\sim$18\% events. Fewest events were recorded in V (0.2\%). It is interesting to note that the five directions (four cardinal N, E, W, S, and V) contain fewer than 9\% of events compared to $>$90\% in the remaining four diagonal directions (NE, SE, SW, SE) resulting in a discontinuous angular distribution. The cause of this discontinuity is not fully understood but one of the key factors is the variation of the muon charge ratio R$_\mu$.

Next, the nine directions are combined into two groups of three directions each; along the horizontal axis shown as N$_{G}$, H$_{G}$, S$_{G}$, and along the vertical axis shown as E$_{G}$, V$_{G}$, W$_{G}$, respectively, in Table\,\ref{Tab_1}. The W$_{G}$ contains only 14\% events compared to 82\% in the E$_{G}$, which is a factor of 6 smaller, displaying a strong east-west asymmetry. However, N$_{G}$, and S$_{G}$ show somewhat comparable statistics at 38\%, and 58\%, respectively, displaying a minor asymmetry of only a factor of 1.5. Consequently, the number of events in the horizontal directions appear to be more symmetrically distributed than the vertical directions. The fraction of events recorded in the V, and W directions are the smallest ($<$1\%). The exact cause of this discontinuous angular distributions is far from understood. However, in the present work the focus is on presentation of the observed data on thunderstorms. An attempt is made to interpret the most basic feature of the east-west asymmetry, namely the ratio of events in these two directions with the aid of CORSIKA simulations.

Clearly, there is no reason for the thunderstorms to occur predominantly in the E$_{G}$ (82\%) than in the W$_{G}$ (18\%) directions. However, a systematically larger value of the muon ratio R$_\mu$\,=\,1.37 in the E$_{G}$ than R$_\mu$\,=\,1.14 in the W$_{G}$ directions as shown in Table\,\ref{Tab_1} implies that a given thunderstorm potential would produce a much larger variation in muon intensity ($\Delta I_{\mu}$) in the E$_{G}$ direction. However, that factor alone can not explain why so few events are observed in the V$_{G}$ direction.

\section{Impact of muon charge asymmetry on thunderstorm event distribution \label{Impact}}

The values of R$_\mu$ obtained with the aid of Monte Carlo simulations using CORSIKA with SIBYLL, and FLUKA generators are shown in Figure\,\ref{Fig_Ratio_Map} for the 169 directions in the FOV of G3MT. For each direction $\sim$10$^6$ muons were simulated, yielding a statistical error of 0.1\%. The details of Monte Carlo simulations are described in Section\,\ref{MC}, which yielded mean R$_\mu$\,=\,1.251 for the full FOV, and has values of 1.136, 1.254, 1.374 for W$_G$, V$_G$, E$_G$ groups, respectively. Similarly, R$_\mu$ values of 1.258, 1.245, 1.261 for N$_G$, H$_G$, S$_G$ groups, respectively are obtained. The mean values of 1.255$\pm$0.119, and 1.255$\pm$0.008 for the N-S, and E-W groups, respectively are nearly the same, yet significant difference in their rms values indicates a stronger muon charge asymmetry in the E-W (rms\,=\,0.119) than the N-S (rms\,=\,0.008) directions. Because most of the thunderstorms develop positive potentials \cite{Stolzenberg_2008,Stough_2022}, which results in deceleration of $\mu^+$ and acceleration of $\mu^-$. This when combined with the fact R$_\mu>$1, (i.e. $\mu^+$ outnumber $\mu^-$) leads to a reduction in the net \imu{}. This decrease in \imu{} is used to estimate the thunderstorm potential as described elsewhere \cite{Hariharan_2019_1,Hariharan_2025}. Since a larger R$_\mu$ results in greater reduction of \imu{}, consequently, the east directions display larger variations in the \imu{}, causing a corresponding increase in the number of thunderstorm events due to this selection effect.

\subsection{Muon charge asymmetry \label{Asymmetry}}

The muon charge ratio R$_\mu\,>\,$1.0 was observed by numerous experiments such as MUTRON \cite{Muraki_1979,Muraki_1983}, MINOS \cite{MINOS_2011,MINOS_2016}, L3+C \cite{L3C}, Bess-TeV \cite{BessTeV}, CosmoALEPH \cite{CosmoALEPH}, OPERA \cite{OPERA}, and CMS \cite{CMS} for energies ranging from hundreds of MeV to over 1\,TeV, that yielded a global mean R$_\mu$\,=\,1.268 above 10\,GeV \cite{World_2001}. However, this value was based on the measurements with different observational conditions that may lead to dependencies on various effects such as the direction (zenith $\theta$, azimuth $\phi$), solar modulation, altitude, and geomagnetic activity, etc. \cite{World_2001}. The geomagnetic effects play a dominant role in the shielding of PCRs, characterized by a direction dependent cutoff rigidity. The PCRs interact in the atmosphere, producing secondary mesons including the pions, and kaons, which rapidly decay into muons, and neutrinos. Because the PCRs are predominantly positively charged, the secondaries contain an excess of positive over negative mesons leading to corresponding excess of positive muons, and an R$_\mu$\,$>$\,1.0. The geomagnetic field causes the cutoff rigidity to vary in the east-west direction, leading to an asymmetry in the surface PCR flux, known as the ``east-west'' effect which was discovered in 1933 \cite{AlvarezCompton_1933,Johnson_1933_1,Johnson_1933_2,Johnson_1935,Johnson_1941,Barber_1949,OgawaNagahara_1950,Rossi_1964}. The PCRs arriving from the west have their paths bent towards the Earth, increasing their intensity, which is quantified by a direction dependent cutoff rigidity of PCRs. The GRAPES-3 FOV has cutoff rigidities ranging from 15 to 24\,GV from west to east directions, respectively that gets reflected in the intensity of muon detected by the G3MT \cite{Hariharan_2019_2,Lipari_2000,Brancusa_2008}. This phenomenon leads to a corresponding asymmetry in R$_\mu$ especially at lower energies ($\lesssim$\,1\,GeV) such as those detected by the G3MT. It is also reflected in a strong dependence on $\theta$ and $\phi$. The muons traversing through increased atmospheric depth at higher $\theta$, display a cos$^2\theta$ dependence \cite{Bahmanabadi_2005,Bahmanabadi_2006}. Another factor contributing to higher values of R$_\mu$ in the east is the increased production of positive pions over negative pions due to higher PCR energies \cite{Rebel_2007}.

\subsection{Monte Carlo simulations \label{MC}}

To investigate if the variation of R$_\mu$ across the G3MT FOV could be simulated, Monte Carlo simulations of particle propagation in the atmosphere were carried out using CORSIKA (\textbf{CO}smic \textbf{R}ay \textbf{SI}mulations for \textbf{KA}scade), a widely used simulation package by the global physics community \cite{corsika}. This package permits the study of EAS development of cosmic gamma-rays, and charged nuclei of different elements in the atmosphere. The secondary particles produced in the interactions of PCRs in the atmosphere are tracked up to a user-defined observational level, and energy thresholds for each secondary particle. At the observational level, the output file stores the position (X, Y, Z), momentum (P$_X$, P$_Y$, P$_Z$), and the time (t) from the first interaction point. The electromagnetic cascades are handled by the electron-gamma shower version-4 (EGS4) \cite{egs4} code, or by the analytical Nishimura-Kamata-Greissen (NKG) function \cite{nkg} as decided by the user. The hadronic interactions in the high-energy domain are treated by one of the external generators that have been incorporated into CORSIKA, which include SIBYLL\,2.3e \cite{sibyll_2.3}, EPOS-LHC-R \cite{eposlhcr}, QGSJETIII-01 \cite{qgsIII}, and DPMJETIII \cite{dpmjet}. Similarly, the low-energy hadronic interactions are handled by one of the two generators, namely, FLUKA\,2024.1 \cite{fluka} or UrQMD\,1.3cr \cite{urqmd}. A combination of high-, and low-energy generators selected by the user is used in the simulations.

\begin{figure*}[t!]
    \centering
        \includegraphics*[width=0.42\textwidth]{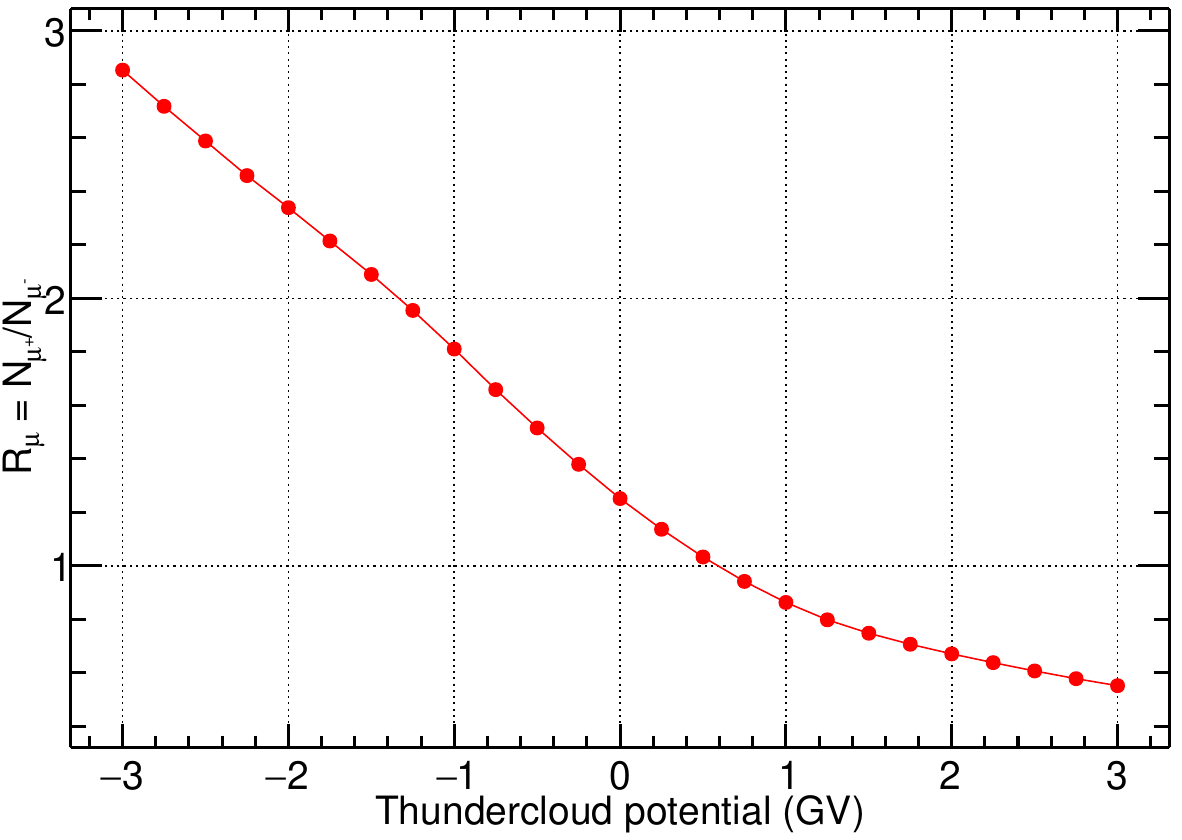}
        \caption{Variation of muon charge ratio (R$_\mu$=N$_{\mu^+}$/N$_{\mu^-}$) as a function of applied thunderstorm potential derived for GRAPES-3 total field of view using CORSIKA with SIBYLL-FLUKA hadronic interaction generators in the thunderstorm potential range of --3\,GV to +3\,GV with a step size of 0.25\,GV. Each potential step contains R$_\mu$ generated using about 169$\times$10$^6$ muons. R$_\mu$ is 1.251 at zero thunderstorm potential.}
        \label{Fig_Ratio_Variation}
    \vspace{0.2cm}
        \includegraphics*[width=0.80\textwidth]{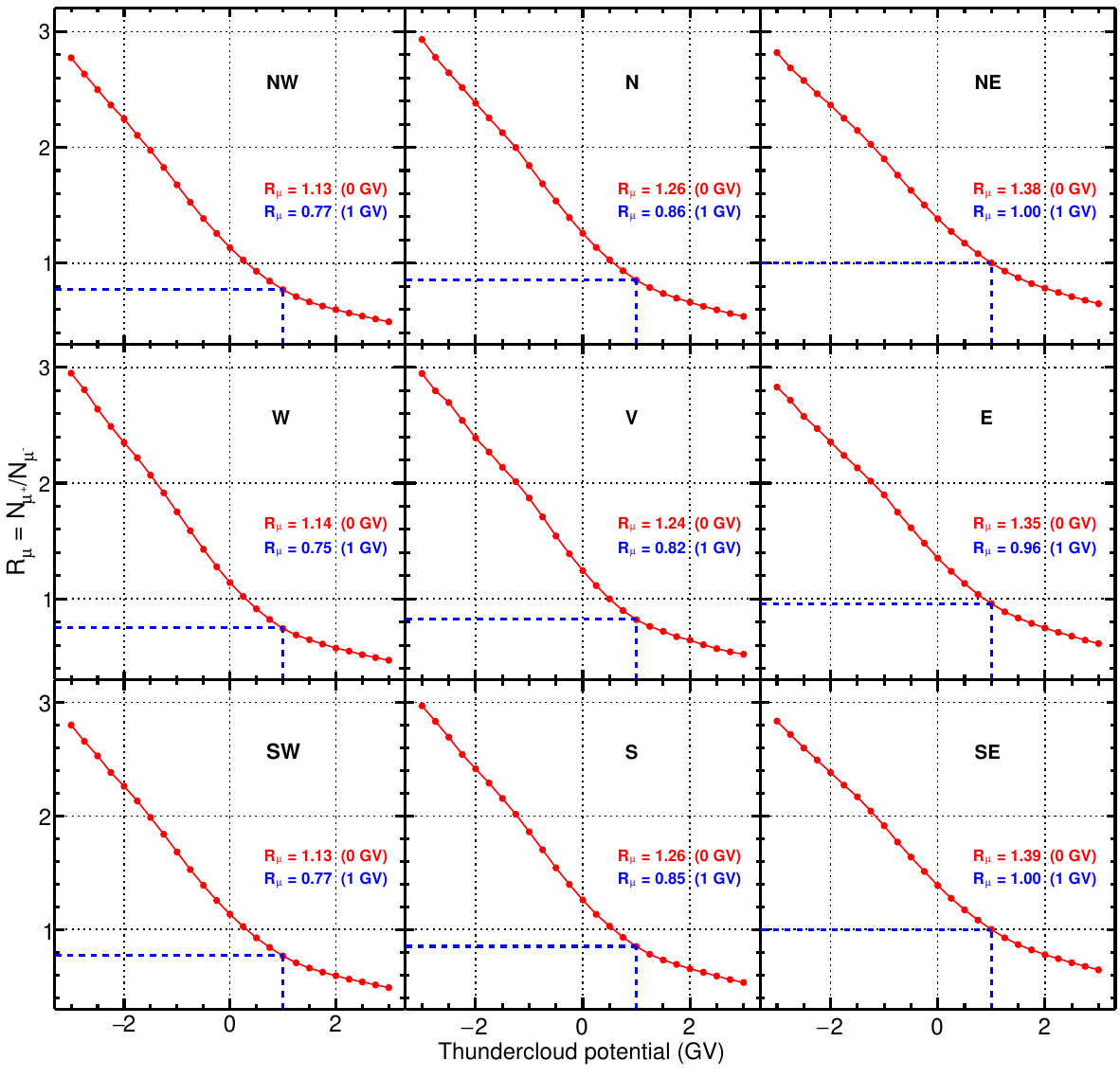}
        \caption{Variation of muon charge ratio R$_\mu$ as a function of applied thunderstorm potential derived for 9 directions in the GRAPES-3 field of view using CORSIKA with SIBYLL-FLUKA generators for a potential range --3\,GV to +3\,GV with step size of 0.25\,GV. The R$_\mu$ values quoted within each subplot are at zero and 1\,GV thunderstorm potentials, respectively.}
        \label{Fig_Ratio_9D}
\end{figure*}

CORSIKA v78000 has been used in this study to simulate protons in the energy range 10\,GeV--10\,TeV. This is because the bulk of the detected muons are produced by the PCRs in this energy range that are predominantly protons ($\sim$90\%). The simulations were carried out for protons with spectral index of --2.65. This index was obtained from a combined fit to the proton energy spectra measured by several direct experiments including PAMELA, CAPRICE, BESS, and CREAM \cite{Chandra_2015}. The efficiency of selecting simulated protons was deliberately enhanced by an option implemented in CORSIKA to reduce the computing time by a factor of $\sim$3 \cite{Hariharan_2015,Hariharan_2019_2}. An in-built atmospheric model ATMOD-5 defined the atmospheric conditions. A modified `EFIELD' option in CORSIKA was implemented for simulating \imu{} as a function of the applied thunderstorm potential \cite{Hariharan_2017_2}. The number of muons detected in the G3MT FOV decline rapidly from the vertical to the inclined directions due to, (a) the steep dependence of muon intensity, (b) the solid angle of the direction bin, on the zenith angle ($\theta$). To compensate for the loss of statistics due to this zenith angle dependence, simulated protons were targeted to the center of each direction instead of distributing them isotropically as is usually done.

During the simulations, the number of PCRs in each of the 169 directions were appropriately scaled to ensure detection of 10$^6$ muons in each direction bin, resulting in a tiny statistical error of only 0.1\%. A wide range of potentials, varying from --3\,GV to +3\,GV in steps of 0.25\,GV were implemented in CORSIKA to simulate the effect of electric potential across a thunderstorm as mentioned above. Each potential was implemented by applying a uniform electric field of required magnitude between two layers from 8 to 10\,km amsl. This choice of thunderstorm altitudes is reasonable, as shown earlier \cite{Hariharan_2019_1,Hariharan_2025}. The bank of 10$^6$ muon in each of the 169 directions was generated and stored for SIBYLL, and FLUKA for the high-, and low-energy hadronic interaction generators, respectively. As discussed in Section\,\ref{Impact}, Figure\,\ref{Fig_Ratio_Map} shows a map of R$_\mu$ in 169 directions at zero thunderstorm potential. R$_\mu$\,=\, 1.251 for the full FOV, and has mean values of 1.136, 1.254, 1.374 for W$_G$, V$_G$, E$_G$ groups, respectively. Similarly, R$_\mu$ has mean values of 1.258, 1.245, 1.261 for N$_G$, H$_G$, S$_G$ groups, respectively. As expected, both groups have the same mean R$_\mu$ of 1.255$\pm$0.119, and 1.255$\pm$0.008 for the horizontal, and vertical groups, respectively, yet, their rms values differ by a factor of $\sim$15 indicating a stronger charge asymmetry for the vertical (rms\,=\,0.119) than for the horizontal groups (rms\,=\,0.008). In Figure\,\ref{Fig_Ratio_Variation} the variation of R$_\mu$ as a function of the applied thunderstorm potential shows a non-linear dependence with gradual flattening at higher potentials. This flattening is a combined outcome of the polarity dependent variation of the muon intensity, and accelerated decays of $\mu^+$ at higher potentials \cite{Hariharan_2025}. In Figure\,\ref{Fig_Ratio_9D}, the variation of R$_\mu$ obtained from CORSIKA is shown as a function of the potential for 9-directions, which also shows a trend similar to Figure\,\ref{Fig_Ratio_Variation}. The dependence on the potential is found to vary from west to east, and systematically higher R$_\mu$ in the east compared to the west turns the east directions more sensitive resulting in larger reductions in \imu{}, exactly as observed in the G3MT data.

\begin{table}[t]
    \centering
    \begin{tabular}{|c|c|c|}
    \hline  
                & FLUKA     & UrQMD     \\
    \hline  
    SIBYLL      & 1.251     & 1.184     \\
                & ---       & (--5.4\%) \\
    \hline  
    EPOS-LHC    & 1.206     & 1.180     \\
                & (--3.6\%) & (--5.7\%) \\
    \hline  
    QGSJETIII   & 1.226     & 1.163     \\
                & (--2.0\%) & (--7.0\%) \\
    \hline  
    DPMJETIII   & ---       & 1.164     \\
                &           & (--7.0\%) \\
    \hline  
    \end{tabular}
    \caption{Muon charge ratio R$_\mu$ derived from CORSIKA with combinations of SIBYLL, EPOS-LHC, QGSJETIII, DPMJETIII for the high-, and FLUKA, UrQMD for low-energy hadronic event generators, respectively. DPMJETIII-FLUKA combination is excluded due to their incompatibility. Values of R$_\mu$ are generated for full G3MT FOV at zero thunderstorm potential, yielding mean R$_\mu$\,=\,1.196$\pm$0.033. R$_\mu$ for the 169 directions is generated by simulating 10$^6$ muon in each direction. SIBYLL-FLUKA combination provides highest mean R$_\mu$\,=\,1.251, thereby providing the most conservative estimate of thunderstorm potential \cite{Hariharan_2025}. UrQMD yields systematically lower R$_\mu$ than FLUKA. QGSJETIII, DPMJETIII yield lowest R$_\mu$ when paired with UrQMD.}
    \label{Tab_2}
\end{table}

\begin{figure}[t]
    \centering
    \includegraphics[scale=0.65]{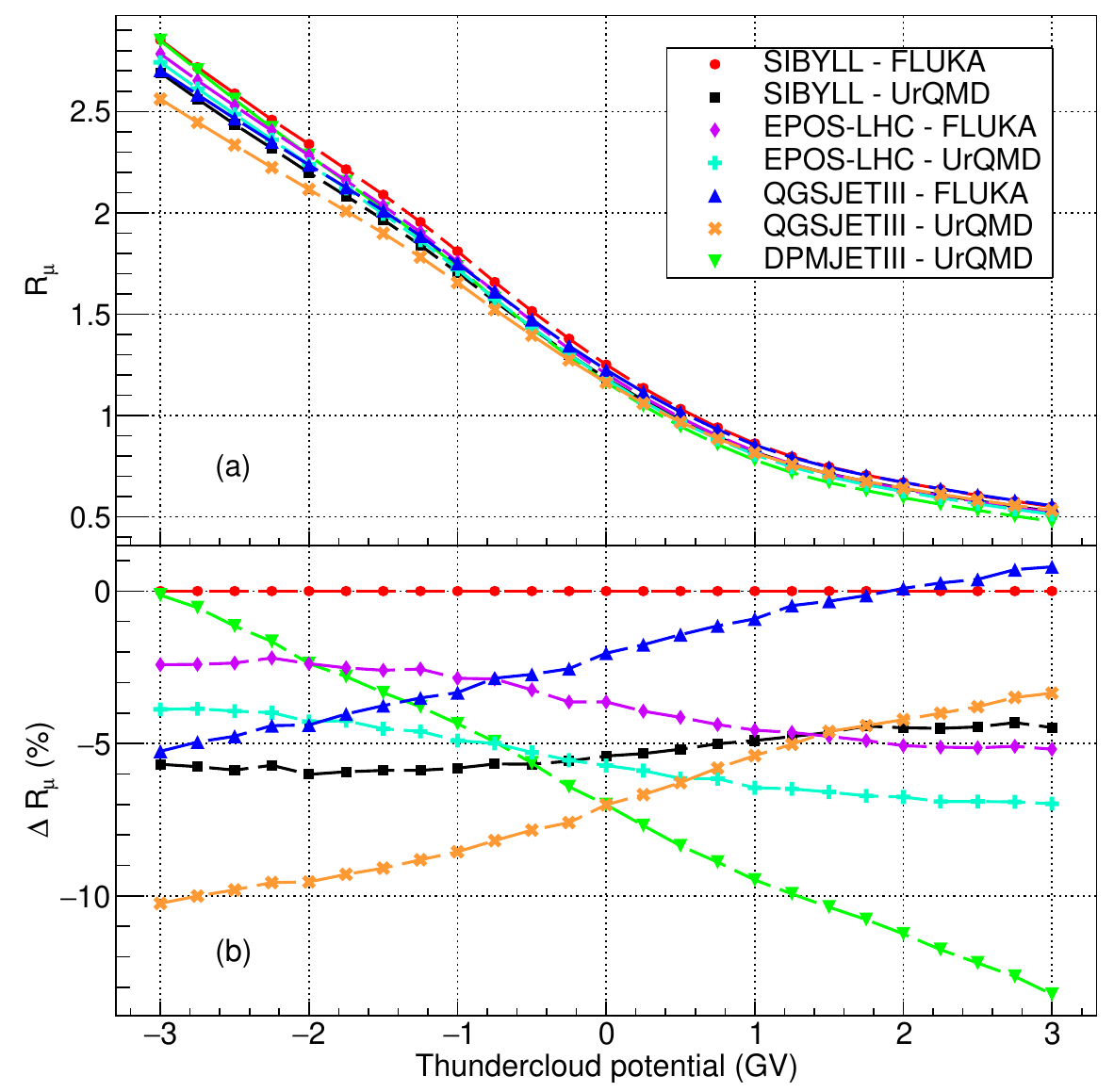}
    \caption{(a) Variation of muon charge ratio R$_\mu$ as a function of applied thunderstorm potential derived for G3MT FOV using CORSIKA for the range of --3\,GV to +3\,GV in step size of 0.25\,GV. The simulations carried out using SIBYLL, EPOS-LHC, QGSJETIII, DPMJETIII for the high-, and FLUKA, UrQMD for low-energy hadronic interaction generators, respectively, (b) Relative variation of R$_\mu$ relative to SIBYLL-FLUKA combination as a function of applied potential. The dotted lines are added to guide the eyes.}
    \label{Fig_Ratio_Models}
\end{figure}

Since these simulations require large computing resources, an effective strategy was devised to create a bank of simulated muons for different combinations of hadronic interaction generators (i.e., EPOS-LHC, QGSJETIII, DPMJETIII for high-, and UrQMD, FLUKA for low-energies, respectively), but with reduced statistics of 10$^5$ muons for each of the 169 directions. It should be noted that the DPMJETIII-FLUKA combination is excluded due to incompatibility. For the remaining seven combinations of generators, R$_\mu$ is shown for the full FOV at zero thunderstorm potential in Table\,\ref{Tab_2} that has a global mean\,=\,1.196$\pm$0.033. SIBYLL-FLUKA combination yielded the highest R$_\mu$\,=\,1.251, which is the primary reason for producing the lowest, and consequently the most conservative estimate of thunderstorm potential for a given change in the muon intensity \cite{Hariharan_2025}. UrQMD yielded systematically lower R$_\mu$ than FLUKA. QGSJETIII and DPMJETIII yielded the lowest R$_\mu$ when paired with UrQMD ($\sim$7\% lower than SIBYLL-FLUKA). Subsequently, the dependence of R$_\mu$ on the hadronic interaction generators is obtained for the entire range of thunderstorm potentials (--3\,GV to +3\,GV), as shown in Figure\,\ref{Fig_Ratio_Models}a for the seven combinations of generators. In Figure\,\ref{Fig_Ratio_Models}b the variation of R$_\mu$ for each of the combinations relative to SIBYLL-FLUKA is shown. As expected, and in agreement with the previous study \cite{Hariharan_2025}, the SIBYLL-FLUKA combination yielded the largest R$_\mu$ for nearly the entire potential range. FLUKA yielded the smallest variation of R$_\mu$ ($\lessapprox$5\%), irrespective of the choice of high-energy generator, whereas UrQMD yielded largest variation of $\sim$14\%. Among the high-energy generators SIBYLL, and EPOS-LHC provided similar responses within a few percent, whereas QGSJETIII, and DPMJETIII provided much higher variations from 6\% to 14\%.

\subsection{Influence on thunderstorm event distribution \label{Influence}}

To investigate the influence of R$_\mu$ on the angular distribution of thunderstorms recorded by G3MT, Monte Carlo simulations were carried out for the three distinct cases. But, first the peak \imu{} values of 487 thunderstorms were converted into equivalent electric potentials using corresponding calibration profiles obtained from SIBYLL-FLUKA based simulations, and their  distribution shown in Figure\,\ref{Fig_Events}. A Gaussian fit to the potential distribution yielded mean\,=\,0.53\,GV, and rms\,=\,0.22\,GV. The positive value of mean potential implies dominant presence of positive potentials in the detected thunderstorms. The location of G3MT in Ooty within the Nilgiris plateau has turned out to be an excellent choice since Ooty experiences nearly round-the-year thunderstorm activity. Next, for each of the nine directions in the FOV, 10$^6$ values of thunderstorm potentials are randomly generated by using a normal distribution with mean\,=\,0.53\,GV, rms\,=\,0.22\,GV. Each potential value is converted back into equivalent \imu{} for that direction with the aid of simulations using CORSIKA with SIBYLL-FLUKA \cite{Hariharan_2025}. Each simulated event is subjected to the cut $\mid\Delta I_{\mu}\mid>$0.3\% exactly as is done for the G3MT data in Section\,\ref{Events}. These data are then combined as W$_G$, and E$_G$ for the west, and east directions, respectively. Their ratio (E$_G$/W$_G$) from the simulation is 5.09$\pm$0.02, which is consistent with 5.9$\pm$0.8 obtained from the G3MT data. It is to be noted that G3MT data contains only 487 thunderstorm events, which contributed to a large statistical error of $\sim$14\% (5.9$\pm$0.8), whereas an arbitrarily large statistics in the simulated data artificially reduced the error (5.09$\pm$0.02).

This exercise reaffirms that the values of R$_\mu\,>\,$1.0 are responsible for enabling the detection of thunderstorms, and the variation of R$_\mu$ across the G3MT FOV causes the observed directional asymmetry among recorded thunderstorms. The east group with R$_\mu$\,=\,1.374, is $\sim$21\% larger than the west group (R$_\mu$=1.136), which led to detection of nearly six times more events in the east. In the second and third iteration, Monte Carlo simulations were carried out to investigate if the east-west asymmetry persisted for a constant value of R$_\mu$ (1.136 and 1.137 for second and third iterations, respectively). These exercises resulted in identical response for both the east, and west groups, implying absence of the east-west effect for the ratio (E$_G$/W$_G$)=\,1.000$\pm$0.002, when kept constant values of R$_\mu$. These simulations supports the earlier conclusion that the east-west asymmetry among thunderstorms is caused by the direction dependent variation of R$_\mu$. The variation of R$_\mu$ in turn is an outcome of the geomagnetic deflection of the PCRs, and their secondaries produced in the atmosphere.

\section{Discussion \label{Discussion}}

The present study demonstrates that the directional asymmetry observed in the distribution of thunderstorm events recorded by the G3MT is primarily caused by the variation of muon charge ratio R$_\mu$, due to the geomagnetic field. The well-known east–west effect of the PCRs gets manifested through the deflection of charged particles in the geomagnetic field, leading to systematically larger values of R$_\mu$ in the east relative to the west directions. This behavior R$_\mu$ is reproduced with the aid of CORSIKA Monte Carlo simulations, including the observation of six times larger number of thunderstorm events in the east compared to the west directions. The simulations have also confirmed that the direction dependent variation of R$_\mu$ in presence of thunderstorm potential results in the observed behavior of \imu{}. The R$_\mu$ values of 1.374, and 1.136 are observed in the east, and west directions accompanied by a east-west ratio of (5.9$\pm$0.8) events. Monte Carlo simulations have reproduced the observed ratio (5.09$\pm$0.02) of thunderstorm events, demonstrating that the directional asymmetry of G3MT to thunderstorms is caused by the geomagnetic field rather than any instrumental bias. 

The dependence of R$_\mu$ on the hadronic interaction generators is also explored. While all combinations of high-, and low-energy generators predict similar qualitative behavior, the absolute value of R$_\mu$ varies by up to $\sim$7\% between different generator sets at zero thunderstorm potential. The SIBYLL-FLUKA pair yields the highest R$_\mu$, and thus the most conservative estimate of thunderstorm potentials. Systematic differences among different generators introduce an uncertainty envelope that should be considered when converting \imu{} to thunderstorm potential. However, the trend remains robust across different combinations, underscoring the fact that the observed asymmetry is a physical phenomenon. It is also noteworthy that the G3MT FOV spans cutoff rigidities from 15 to 24\,GV, offering a sizable variation in the geomagnetic bending that illustrates the east–west asymmetry. The high altitude Nilgiris plateau where Ooty is located is subjected to frequent convective activity, providing favorable conditions for generating strong thunderstorms with potentials, often exceeding 1\,GV \cite{Hariharan_2019_1,Hariharan_2025}. The combination of geomagnetic filtering, and local meteorological factors makes G3MT particularly sensitive to thunderstorm-induced activity. The close agreement between the observed data, and simulations implies that muon charge asymmetry is the dominant driver for the highly asymmetric distribution of thunderstorm events. These findings not only refine our understanding of the estimates of potential difference across thunderstorms, but also strengthen the role of G3MT as a powerful probe of the atmospheric electrodynamics.

\section{Conclusions \label{Conclusions}}

In the present study 487 thunderstorm events recorded by the G3MT between April 2011 and December 2020 display a strong asymmetry in their distribution with six times (5.9$\pm$0.8) more events observed in the east than west directions. Monte Carlo simulations using CORSIKA show this behavior arose from the muon charge asymmetry caused by the geomagnetic field on the PCRs and their secondaries during transport in the interplanetary space, and in the atmosphere, respectively. The simulations show increase in muon charge ratio R$_\mu$ (N$_{\mu^+}/N_{\mu^-}$) from the west to east directions that reproduced the variation of muon intensity during thunderstorms, and of dominance of thunderstorms from east. Simulations reproduced the ratio of east-west events (5.09$\pm$0.02). Among different hadronic interaction generators, SIBYLL–FLUKA combination yielded largest R$_\mu$, and therefore, the most conservative estimate of thunderstorm potentials. Despite a relatively small generator dependent variation, this trend is consistent across all seven generator combinations. In summary, R$_\mu>$1 essentially made the detection of thunderstorm by the G3MT possible, and its variation is responsible for the asymmetric thunderstorm distribution in the G3MT. These findings strengthen the physical basis for using the muon observations to study massive thunderstorm potentials, and their interactions with the geomagnetic field.

\section*{Acknowledgement}

The authors thank S. Kingston, K. Manjunath, S. Murugapandian, S. Pandurangan, B. Rajesh, V. Santhoshkumar, M.S. Shareef, C. Shobana and R. Sureshkumar for their efforts in maintaining the GRAPES-3 experiment. The authors also express sincere thanks to the remaining members of the collaboration for their reviews and comments. We acknowledge the support of the Department of Atomic Energy, Government of India, under Project Identification No. RTI4002.


\end{document}